\documentclass[referee]{raa}            

\usepackage{graphicx,times}             
\usepackage{natbib}
\usepackage{amssymb,amsmath}
\usepackage{rotating}
\bibpunct{(}{)}{;}{a}{}{,}

\usepackage[pagebackref=true]{hyperref}

\begin{document}

  \title{WFST Supernovae in the First Year: I. Statistical Study of 16 Early-phase Type Ia Supernovae from the Pilot Survey}

   \volnopage{Vol.0 (20xx) No.0, 000--000}      
   \setcounter{page}{1}          

   \author{Weiyu Wu 
      \inst{1,2}
   \and Ji-an Jiang
      \inst{1,3,*}
   \and Zelin Xu
      \inst{1,2}
   \and Dezheng Meng
      \inst{1,2}
   \and Keiichi Maeda
      \inst{4}
   \and Hanindyo Kuncarayakti
      \inst{5,6}
   \and Lluís Galbany
      \inst{7,8}
   \and Saurabh W. Jha
      \inst{9}
   \and \v{Z}eljko Ivezi\'{c}
      \inst{10}
   \and Peter Yoachim
      \inst{10}
   \and Zhengyan Liu
      \inst{1}
   \and Junhan Zhao
      \inst{1,2}
   \and Tinggui Wang
      \inst{1,11}
   \and Xu Kong
      \inst{1,11}
   \and Andrew J. Connolly
      \inst{10}
   \and Ziqing Jia
      \inst{1,2}
   \and Lei Hu
      \inst{12}
   \and Lulu Fan
      \inst{1,11}
   \and Ning Jiang
      \inst{1}
   \and Feng Li
      \inst{13}
   \and Ming Liang
      \inst{14}
   \and Jinlong Tang
      \inst{15}
   \and Zhen Wan
      \inst{1}
   \and Hairen Wang
      \inst{16}
   \and Jian Wang
      \inst{11,13}
   \and Yongquan Xue
      \inst{1}
   \and Hongfei Zhang
      \inst{13}
   \and Wen Zhao
      \inst{1}
   \and Xianzhong Zheng
      \inst{17}
   \and Qingfeng Zhu
      \inst{1,11}
   }


   \institute{Department of Astronomy, University of Science and Technology of China, Hefei 230026, China; {\it (Corresponding author) \it jian.jiang@ustc.edu.cn}\\
        \and
             School of Astronomy and Space Sciences, University of Science and Technology of China, Hefei, 230026, China\\
        \and
             National Astronomical Observatory of Japan, 2-21-1 Osawa, Mitaka, Tokyo 181-8588, Japan\\
        \and
             Department of Astronomy, Kyoto University, Kitashirakawa-Oiwake-cho, Sakyo-ku, Kyoto 606-8502, Japan\\
        \and
             Tuorla Observatory, Department of Physics and Astronomy, FI-20014 University of Turku, Finland\\        
        \and
             Finnish Centre for Astronomy with ESO (FINCA), FI-20014 University of Turku, Finland\\
        \and
             Institute of Space Sciences (ICE-CSIC), Campus UAB, Carrer de Can Magrans, s/n, E-08193 Barcelona, Spain\\
        \and
             Institut d'Estudis Espacials de Catalunya (IEEC), 08860 Castelldefels (Barcelona), Spain\\
        \and
             Department of Physics and Astronomy, Rutgers, The State University of New Jersey, 136 Frelinghuysen Road, Piscataway, New Jersey 08854, USA\\
        \and
             Department of Astronomy, University of Washington, Box 351580, Seattle, Washington 98195-1580, USA\\
        \and
             Institute of Deep Space Sciences, Deep Space Exploration Laboratory, Hefei 230026, China\\
        \and
             McWilliams Center for Cosmology, Department of Physics, Carnegie Mellon University, 5000 Forbes Ave, Pittsburgh, 15213, PA, USA\\
        \and
             State Key Laboratory of Particle Detection and Electronics, University of Science and Technology of China, Hefei 230026, China\\
        \and
             National Optical Astronomy Observatory (NSF’s National Optical-Infrared Astronomy Research Laboratory) 950 N Cherry Ave. Tucson Arizona 85726, USA\\
        \and
             Institute of Optics and Electronics, Chinese Academy of Sciences, Chengdu 610209, China\\
        \and
             Purple Mountain Observatory, Chinese Academy of Sciences, Nanjing 210023, China\\
        \and
             Tsung-Dao Lee Institute and Key Laboratory for Particle Physics, Astrophysics and Cosmology, Ministry of Education, Shanghai Jiao Tong University, Shanghai, 201210, China\\
\vs\no
   {\small Received 20xx month day; accepted 20xx month day}}

\abstract{ In this paper we present 16 early-phase type Ia supernovae (SNe Ia) discovered during the pilot survey of the 2.5-meter Wide Field Survey Telescope (``WFST-PS") from March 4 to July 10, 2024, including three SNe Ia with early-excess emission features (EExSNe Ia). The discovery magnitude of the 16 WFST-PS early-phase SNe is at least 3 mag fainter than their peak brightness. A large scatter of color indices is found in approximately the first 10 days of supernova explosions, indicating diverse photometric behaviors in the early phase. Three EExSNe Ia show relatively brighter peak luminosities and longer rise time compared to those of non-EExSNe Ia. The results indicate that current theoretical models require further refinement to fully capture the early photometric evolution of SNe Ia. Based on the initial high-cadence \(ugr\)-band data from the WFST-PS survey, we emphasize that early near-ultraviolet (NUV) observations are indispensable for placing tight constraints on the explosion mechanisms and progenitor systems of SNe Ia.
}

   \authorrunning{W.-Y. Wu}            
   \titlerunning{WFST-PS Early-phase SNe Ia}  

   \maketitle

%
%
\section{Introduction}           
\label{sect:intro}

The prevailing view in the academic community holds that type Ia supernovae (SNe Ia) originate from thermonuclear explosions of carbon-oxygen white dwarfs. These explosions are triggered either by the continuous accretion of material from a non-degenerate companion star (the single-degenerate channel, {\citealt{1973ApJ...186.1007W}) or through a merger with a companion star (the double-degenerate channel, \citealt{1984ApJ...284..719I,1984ApJ...279..252K}). The homogeneous peak luminosity of normal SNe Ia ($\sim 70\%$ of the total population;\citealt{2012AJ....143..126B}) enables their use as precise cosmological standard candles, following calibration via the empirically derived Phillips relation (light-curve shape vs. peak brightness; \citealt{1993ApJ...413L.105P,1996AJ....112.2391H,1997ApJ...483..565P,1998AJ....116.1009R,1999AJ....118.1766P,1999ApJ...517..565P}). However, significant disagreement persists regarding the nature of the progenitor binary systems and the precise explosion mechanisms, thereby compromising the accuracy of normal SNe Ia as cosmological distance indicators. The recent surge in transient surveys has pioneered new avenues for studying the properties of transients within days of explosion \citep{2022ApJ...933L..36J,2025JCAP...08..053G}. Increasingly, the ultra-early light-curve behavior of normal SNe Ia has been observed and utilized to constrain specific explosion mechanisms, offering fresh insights into their progenitor systems and explosion physics \citep{kasen2009seeing,10.1093/pasj/psv028,jiang2017hybrid,maeda2018type}.

During the early light-curve rise phase, observations have revealed diverse light-curve morphologies\citep{firth2015rising,2019MNRAS.483.5045P,2022MNRAS.514.3541S,2023ApJ...956L..34S,2024ApJ...962...17W}. Furthermore, \cite{jiang2018surface} demonstrated through a large sample analysis that SNe Ia show diverse flux excess light-curve morphologies during their early phases. \cite{kasen2010optical} first studied how progenitor systems affect the early light-curves of SNe Ia. The study found that when the explosion ejecta hits a companion star, it creates ultraviolet and optical light for several days. If observed from certain directions, this interaction causes a clear brightening in the first few days after the explosion \citep{2021ApJ...923L...8J}. \cite{2016ApJ...826...96P} studied the ejecta-circumstellar material interaction, which produces early-phase $UV$-optical excess flux similar to companion interactions. Clumps of $^{56}\text{Ni}$ in the outer ejecta may create early flux excess before the main light-curve rise dominated by inner $^{56}\text{Ni}$ \citep{2019ApJ...870L..14D,2020A&A...642A.189M}. The observed rising-phase morphologies strongly depend on the spatial distribution of $^{56}\text{Ni}$ within the ejecta \citep{2014ApJ...784...85P,2016ApJ...826...96P,2017MNRAS.472.2787N,2018A&A...614A.115M}. Additionally, variations in the mass and composition of helium shells can produce diverse early phase light-curve behaviors \citep{maeda2018type,2021MNRAS.502.3533M}. In the double-detonation model, the ignition of a helium shell on a white dwarf synthesizes short-lived radioactive isotopes (e.g., $^{56}\text{Ni}$, $^{52}\text{Fe}$, $^{48}\text{Cr}$). Their decay powers early phase emission within days of the explosion \citep{jiang2017hybrid,2017MNRAS.472.2787N,2019ApJ...873...84P,2021MNRAS.502.3533M}. \cite{2021MNRAS.502.3533M} shows that models with helium shells lacking iron-group elements reproduce normal SNe Ia across a range of luminosities, matching observations from explosion through peak brightness. \cite{2021ApJ...922...68S} shows that sub-Chandrasekhar double-detonation models under varying viewing angles broadly match the light-curves and spectra of normal SNe Ia across all luminosities, from sub- to super-luminous.

In early phases, in addition to the overall light-curve morphology, color evolution can effectively probe the spatial distribution of radioactive isotopes such as $^{56}\text{Ni}$ in the ejecta \citep{10.1093/mnras/stu598} and serves as an important basis for distinguishing different theoretical models \citep{2023RAA....23h2001L}. For the double-detonation model, radioactive elements are produced in the outermost and inner regions, leading to the formation of a unique red bump feature \citep{2017MNRAS.472.2787N,maeda2018type,2019ApJ...873...84P}. However, optical photometry of early phase light-curve features alone may not suffice to discriminate helium-detonation models from alternative explosion mechanisms. When combined with ultraviolet observations, the identification of early phase behaviors becomes possible. Theoretically,  iron-peak elements strongly influence ultraviolet radiation through line-blanketing and line-blocking effects \citep{1992ApJ...397..304J,2008MNRAS.391.1605S,2013MNRAS.429.2228H}. Models show that the ultraviolet band is extremely sensitive to the nucleosynthetic products (i.e., ejecta composition) of supernova explosions \citep{2008MNRAS.391.1605S,2000ApJ...530..966L}. Therefore, early ultraviolet observations can sensitively detect the composition of the outer ejecta material. Helium-detonation models predict a $UV$ peak within $~1$ day post-explosion, significantly earlier than interaction models, and show redder $UV$-optical colors in the initial days \citep{2021MNRAS.502.3533M}. Furthermore, \citet{2013ApJ...779...23M} discovered two distinct subgroups of normal SNe Ia based on their $NUV$-optical colors, utilizing a large dataset from the Swift UV/Optical Telescope (UVOT).

Here we report 16 SNe Ia with multiband early observations or deep non-detection constraints discovered during the pilot survey of Wide Field Survey Telescope (WFST\footnote{https://wfst.ustc.edu.cn}), three of them show clear early-excess features. The paper is organized as follows. The WFST transient survey design and sample selection criteria are outlined in Section~\ref{sec:wfst}. Photometric behavior and basic analyses of early-phase SNe Ia are summarized in Section~\ref{sec:sample} and Section~\ref{sec:res}. Further discussions and conclusions are given in Section~\ref{sec:dis}.


\section{Observations and Sample Overview} \label{sec:wfst}
The WFST is a 2.5-meter optical telescope jointly built by the University of Science and Technology of China (USTC) and the Purple Mountain Observatory (PMO), which is designed for a high throughput in blue wavelengths and equipped with a mosaic CCD camera (9 × 9k × 9k) with a field of view of  $\sim6.5$ square degrees \citep{Wang2023}. 

The pilot survey of WFST (``WFST-PS") was carried out from Mar 4th to July 10, 2024, which aims to evaluate the performance of transient search \citep{2022Univ....9....7H} and carry out early sciences with WFST. WFST started the 6-year transient survey from Dec 14, 2024. All SNe shown in this article were discovered by the Deep High-cadence $ugr$-band Survey project (``DH$ugr$"), a key project for both pilot and formal surveys. The primary observing strategy involved daily/hourly-cadence photometries in $u$, $g$, and $r$ bands\footnote{However, due to an unexpected technical issue of the filter-exchange system occurred in late Mar 2024, only $g$- and $r$-bands data were obtained during the remaining WFST-PS period.}. Standard data processing was implemented through the WFST data pipeline, which is built on a modified version of the Large Synoptic Survey Telescope (LSST) software stack \citep{2010SPIE.7740E..15A,2018PASJ...70S...5B,2019ApJ...873..111I}. More details on WFST data processing can be found in \cite{2025arXiv250115018C} and Z. Xu et al. (2025, in prep). 

Team members carried out visual inspections of all WFST-PS transient candidates after the real-bogus classifier. Based on specific criteria (incl host-center offset, redshift, light-curve morphology, source catalog cross-match, etc.), we ultimately classified 894 sources as robust SN candidates. Refer J. Jiang et al. (2025, in prep) for further details of the DH$ugr$ project and the whole SN sample discovered by WFST-PS.

\subsection{SN Ia Classification} \label{subsec:selection}
Superphot+ is a photometric classifier that employs parametric modeling to extract meaningful features from multi-band supernova light-curves without requiring redshift information, while maintaining classification accuracy comparable to redshift-dependent methods \citep{2024ApJ...974..169D}. Using Superphot+, we further classified our sample of 894 confirmed SNe, identifying 411 SNe Ia, of which 363 have either host galaxy spectroscopic redshifts or photometric redshift. We identified 16 SNe Ia as WFST early-phase SNe Ia based on the criterion that their detection magnitude was at least 3 magnitudes fainter than peak brightness in single or multiple bands. For our analysis, we used exclusively the observed light-curves after time-dilation correction and directly compared them with the fit results to avoid introducing additional uncertainties from applying K-corrections to the early-phase.

For 16 early-phase SNe Ia, we did follow-up spectroscopic observations using the Double Spectrograph (DBSP; \citealt{1982PASP...94..586O}) mounted on the Palomar 200-inch (P200) telescope. However, due to unexpected weather conditions and technical issues, we were unable to obtain usable spectra for any target. Spectroscopic identifications were obtained for four early-phase SNe Ia, WFST-PS240407h (2024gmn), WFST-PS240410d (2024gme), WFST-PS240504x (2024kxr), and WFST-PS240513c (2024jbb) from the Transient Name Server (TNS). The Supernova Identification tool (SNID; \citealt{2007ApJ...666.1024B}) employs cross-validated template matching to determine supernova subtypes. For SNe Ia, SNID classifies them into five distinct subtypes: Ia-norm (normal), Ia-91T (91T-like), Ia-91bg (91bg-like), Ia-csm (circumstellar material interacting), and Ia-pec (peculiar). Our analysis utilized an expanded version of the SNID template library (v2.0), which incorporates not only the original spectral templates but also includes templates for tidal disruption events (TDEs) and superluminous supernovae (SLSNe). We applied SNID to classify the spectral subtypes of our four early-phase SNe Ia. The 16 early-phase SNe Ia span a redshift range of $0.0181 < z < 0.165$ (see Figure~\ref{fig:z}), with redshifts derived from either spectroscopic or photometric measurements of their host galaxies. To ensure the reliability of the classification results and minimize the impact of noise or data quality on the classification, we require the relative likelihood a posteriori ($rlap$) to be $\geq 5$. The template matching results were visually verified using PySNID\footnote{https://github.com/MickaelRigault/pysnid}, as shown in Figure~\ref{fig:spec_classify}. Through this comprehensive analysis, we confirmed that three early-phase SNe Ia (WFST-PS240407h, WFST-PS240410d, WFST-PS240513c) belong to the normal type (Ia-norm) and WFTS-PS240504x belongs to 91T-like. To validate the reliability of the classification, we performed light-curve fitting using the SALT2 model \citep{2010A&A...523A...7G,2014A&A...568A..22B} on the remaining 12 early-phase SNe Ia. All SNe were fitted with snfit (v2.4)\footnote{http://supernovae.in2p3.fr/salt/doku.php?id=usage} using photometric data in  $g$, $r$, and $u$ bands when available. A more detailed discussion of these results are presented in Section~\ref{subsec:lc}.

\begin{figure*}
    \centering
    \includegraphics[width=0.7\textwidth]{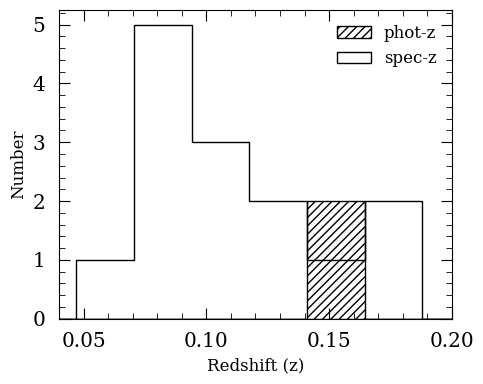}
    \caption{The redshift distribution of WFST-PS early-phase SNe Ia. Photometric redshifts of two host galaxies are used due to the missing of spectroscopic redshifts of both SNe and hosts.
    \label{fig:z}}
\end{figure*}

\begin{figure*}
    \centering
    \includegraphics[width=0.8\textwidth]{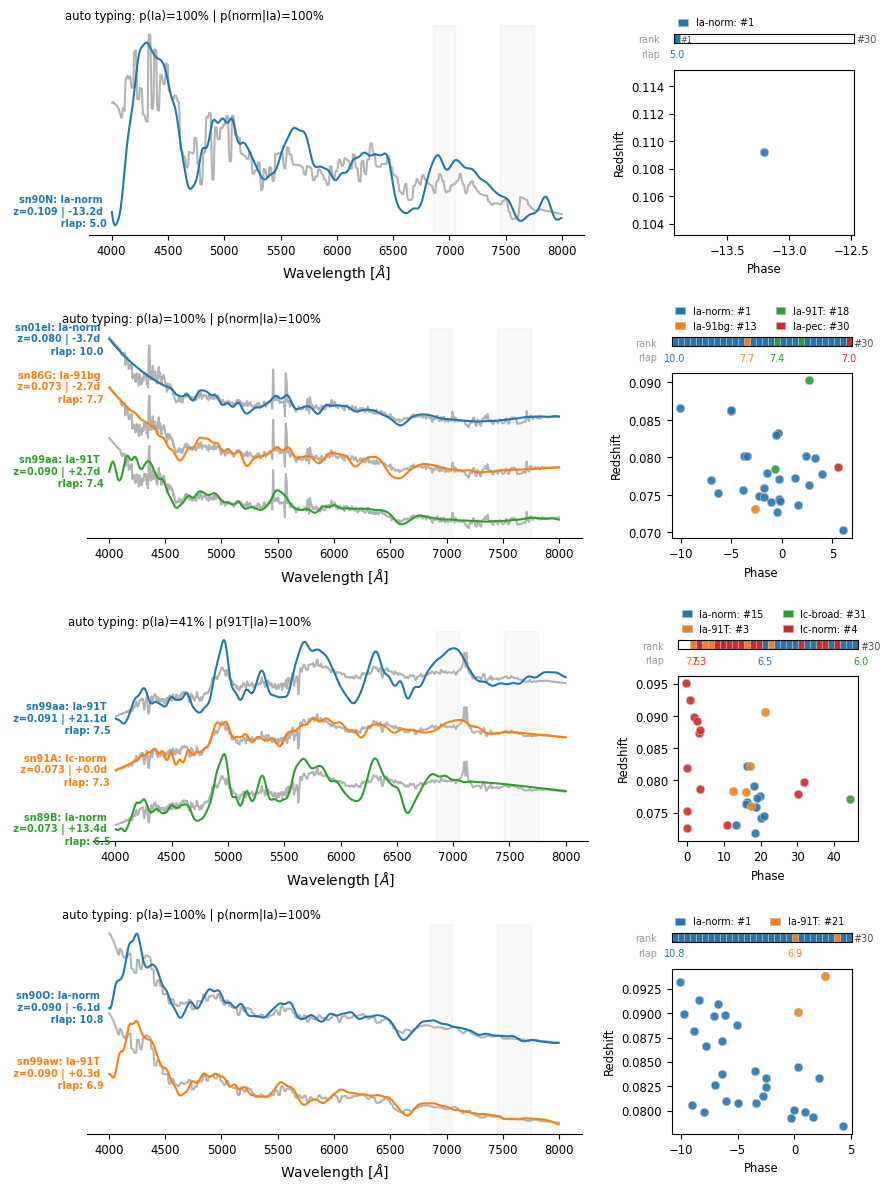}
    \caption{Spectra and SNID classification results for four early-phase SNe Ia. From top to bottom: WFST-PS240407h, WFST-PS240410d, WFST-PS240504x, and WFST-PS240513c. The colored lines and their corresponding labels represent the matched template spectra, while the gray lines show the observed spectra of each early-phase SNe Ia. The probability matches with $rlap \geq 5$ are presented to ensure reliable classification.}
    \label{fig:spec_classify}
\end{figure*}
\begin{figure*}
    \centering
    \includegraphics[width=1\textwidth]{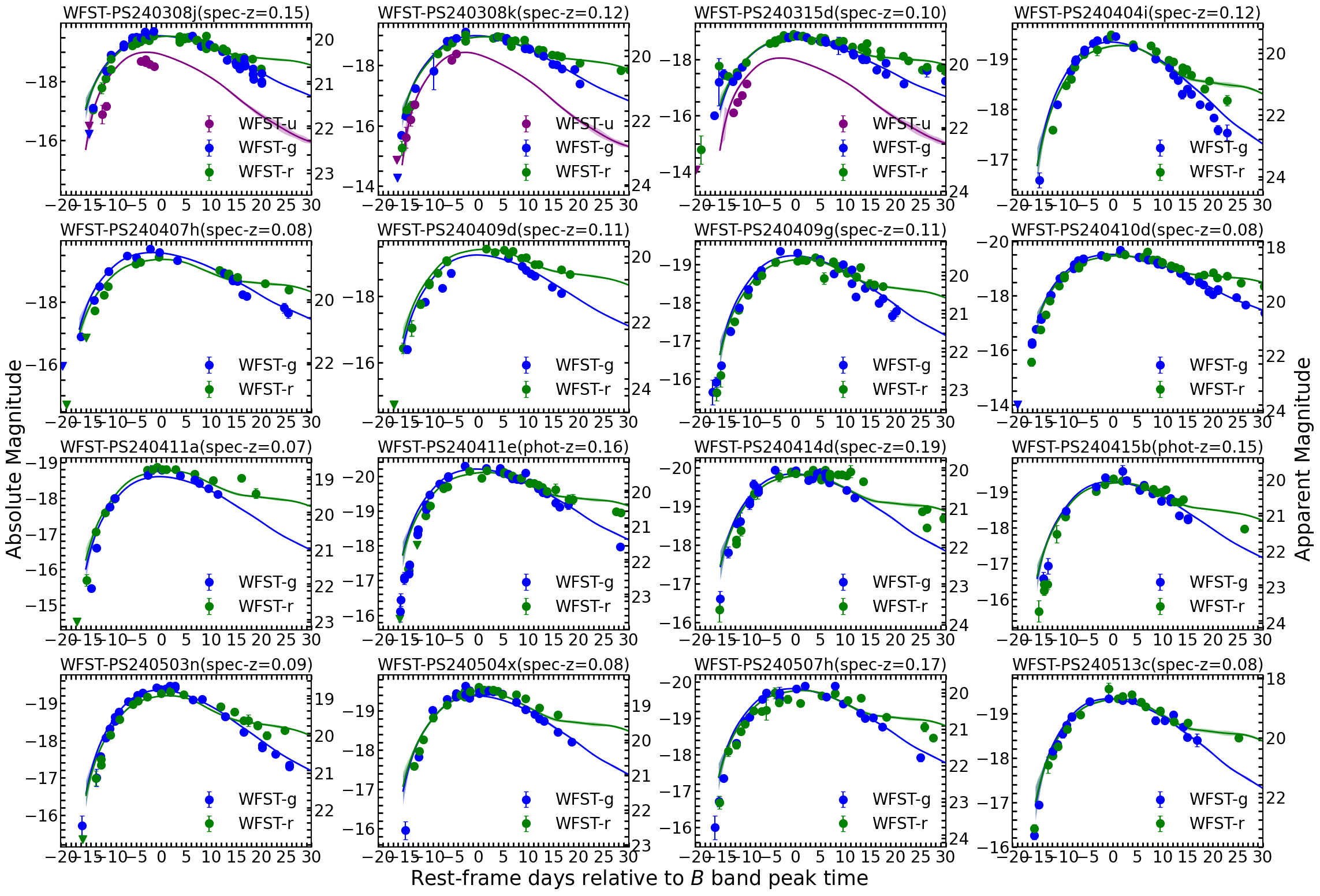}
    \caption{Light curves of 16 WFST-PS early-phase SNe Ia in the $u$, $g$, and $r$ bands with their SALT2 best-fit models. The $u$, $g$, and $r$ bands are represented by purple, blue, and green colors, respectively.
    \label{fig:lc-salt2-fit}}
\end{figure*}
\section{WFST-PS SN\lowercase{e} I\lowercase{a}} \label{sec:sample}
\subsection{Light Curve} \label{subsec:lc}
To estimate the time of $B$-band maximum ($t_{B,\text{max}}$), the observed peak magnitude ($m_{B,\text{max}}$), and the \(B\)-band decline rate parameter, \(\Delta m_{15}(B)\)}, we performed SALT2 light-curve fitting for 16 early-phase SNe Ia. SALT2 parameterizes the flux density of a supernova as a function of phase and rest-frame wavelength, where $x_0$, $x_1$, and $c$ represent the normalization, shape, and color parameters, respectively. Table~\ref{tab:salt2} presents the fit parameters of 16 WFST-PS early-phase SNe Ia. The measured parameter ranges ($-0.13 < c < 0.21$ and $0.22 < x_1 < 1.40$) are consistent with criteria of cosmology-used normal SNe Ia ($-0.3 \leq c \leq 0.3$, $-3.0 \leq x_1 \leq 3.0$; \citealt{2022ApJ...938..110B}). 

\begin{table}
\begin{center}
\caption[]{Light-curve fitting parameters of WFST-PS early-phase SNe Ia\label{tab:salt2}}
 \begin{tabular}{llllll}
  \hline\noalign{\smallskip}
Name & $\Delta m$15(B) & ${x_0}$ & ${x_1}$ & Color & Stretch\\
  \hline\noalign{\smallskip}
WFST-PS240308j & 0.97489(0.08951) & 0.00020(0.00000) & 0.74957(0.09048) & 0.04534(0.02062) & 1.04958(0.09048)\\
WFST-PS240308k & 1.04390(0.06715) & 0.00034(0.00001) & 0.30704(0.06881) & 0.02388(0.02236) & 1.00820(0.06881)\\
WFST-PS240315d & 0.97135(0.05129) & 0.00037(0.00001) & 0.79219(0.05144) & 0.11937(0.02277) & 1.05360(0.05144)\\
WFST-PS240404i & 0.96097(0.12363) & 0.00031(0.00001) & 0.85550(0.12363) & 0.00286(0.02925) & 1.05958(0.12363)\\
WFST-PS240407h & 1.06837(0.07605) & 0.00085(0.00003) & 0.88448(0.09118) & -0.13024(0.02913) & 1.06232(0.09118)\\
WFST-PS240409d & 1.05119(0.18568) & 0.00019(0.00001) & 0.22176(0.18900) & 0.20445(0.03708) & 1.00032(0.18900)\\
WFST-PS240409g & 0.99393(0.11226) & 0.00031(0.00001) & 0.62642(0.11226) & -0.02129(0.02900) & 1.03800(0.11226)\\
WFST-PS240410d & 0.97519(0.05273) & 0.00094(0.00003) & 0.78711(0.05363) & 0.02092(0.02732) & 1.05312(0.05363)\\
WFST-PS240411a & 1.00334(0.10457) & 0.00055(0.00002) & 0.58001(0.10446) & 0.18334(0.02873) & 1.03364(0.10446)\\
WFST-PS240411e & 0.95562(0.09063) & 0.00037(0.00001) & 0.89125(0.09101) & -0.02382(0.02821) & 1.06296(0.09101)\\
WFST-PS240414d & 0.93618(0.13867) & 0.00019(0.00001) & 1.03107(0.13869) & 0.02117(0.02871) & 1.07619(0.13869)\\
WFST-PS240415b & 1.03638(0.12961) & 0.00019(0.00001) & 0.35429(0.13283) & 0.02175(0.02995) & 1.01258(0.13283)\\
WFST-PS240503n & 1.06476(0.06269) & 0.00063(0.00002) & 0.23045(0.07682) & -0.07882(0.02828) & 1.00112(0.07682)\\
WFST-PS240504x & 0.94940(0.09189) & 0.00074(0.00002) & 0.93761(0.09189) & 0.08241(0.02784) & 1.06734(0.09189)\\
WFST-PS240507h & 0.88333(0.16929) & 0.00022(0.00001) & 0.91899(0.16447) & 0.01182(0.03106) & 1.06558(0.16447)\\
WFST-PS240513c & 0.87741(0.15079) & 0.00072(0.00003) & 1.04259(0.18527) & 0.06952(0.03236) & 1.07729(0.18527)\\
  \noalign{\smallskip}\hline
\end{tabular}
\end{center}
\end{table}

The photometry characteristics of 16 WFST-PS early-phase SNe Ia are presented in Table~\ref{tab:sample}. We follow WFST-PS names for transients. Some events were first reported by other groups, but the WFST forced photometry shows detections earlier than the TNS reports. Absolute magnitudes were determined from the host spectroscopic or photometric redshift and Galactic extinction $E(B-V)$, with extinction estimates as refined by \cite{2011ApJ...737..103S}, adopting $R_V = 3.1$. No host extinction corrections were applied.

\begin{table}
\bc
\begin{minipage}[]{100mm}
\caption[]{Characteristics of WFST-PS early-phase SNe Ia}\label{tab:sample}\end{minipage}
\setlength{\tabcolsep}{0pt}
\small
 \begin{tabular}{ccccccccccccc}
  \hline\noalign{\smallskip}
WFST-PS Name & TNS Name & EEx & $\Delta{t_{m(-13)}}$ & R.A. & Dec & Redshift$^a$ & E$(B-V)_{MW}$ & $T_{Bmax}$ & $B_{peak}$ & WFST Discovery$^b$\\
  \hline\noalign{\smallskip}
WFST-PS240308j & 2024mcq & N & 17.17(0.01) & 13:21:22.46 & +01:15:04.09 & 0.15(0.03) & 0.03 & 60392.53(0.07) & -19.66(0.02) & Y\\
WFST-PS240308k & 2024fet & N & 16.38(0.02) & 10:15:43.30 & +04:40:56.80 & 0.12(0.03) & 0.03 & 60393.46(0.06) & -19.07(0.02) & Y\\
WFST-PS240315d & 2024fix & Y & 18.50(0.10) & 13:35:45.10 & +05:19:39.12 & 0.101(0.02) & 0.03 & 60401.33(0.06) & -19.50(0.02) & Y\\
WFST-PS240404i & 2024gko & N & 16.70(0.10) & 13:27:20.39 & +02:18:34.15 & 0.12(0.03) & 0.03 & 60420.08(0.07) & -19.45(0.03) & Y\\
WFST-PS240407h & 2024gmn & Y & 18.90(0.10) & 14:32:12.28 & +04:59:33.20 & 0.09(0.00) & 0.04 & 60421.21(0.05) & -19.60(0.03) & Y\\
WFST-PS240409d & 2024nqg & N & 17.60(0.05) & 11:32:47.39 & -2:24:44.90 & 0.11(0.00) & 0.03 & 60422.80(0.12) & -19.26(0.04) & Y\\
WFST-PS240409g & 2024gsq & N & 17.58(0.20) & 13:09:09.78 & +04:54:27.81 & 0.11(0.00) & 0.03 & 60424.23(0.07) & -19.33(0.03) & Y\\
WFST-PS240410d & 2024gme & N & 17.40(0.10) & 14:09:42.74 & +03:58:15.24 & 0.08(0.00) & 0.03 & 60427.13(0.04) & -19.61(0.03) & Y\\
WFST-PS240411a & 2024hjz & N & 16.44(0.02) & 10:18:28.58 & -1:55:42.71 & 0.07(0.00) & 0.05 & 60425.51(0.08) & -19.08(0.03) & N\\
WFST-PS240411e & 2024wis & N & 17.53(0.01) & 14:27:21.34 & +02:15:43.98 & 0.16(0.00)$^*$ & 0.03 & 60427.84(0.07) & -19.34(0.03) & Y\\
WFST-PS240414d & 2024lvg & N & 17.40(0.01) & 14:25:09.89 & -1:11:35.07 & 0.19(0.00) & 0.05 & 60429.65(0.14) & -19.07(0.03) & Y\\
WFST-PS240415b & 2024icu & N & 15.27(0.02) & 10:54:18.98 & +01:24:01.17 & 0.15(0.00)$^*$ & 0.03 & 60426.42(0.10) & -19.43(0.03) & Y\\
WFST-PS240503n & 2024ils & N & 17.60(0.20) & 13:32:53.31 & -1:27:34.97 & 0.09(0.00) & 0.03 & 60446.78(0.05) & -19.52(0.03) & Y\\
WFST-PS240504x & 2024kxr & N & 15.10(0.08) & 14:49:02.04 & -2:13:08.77 & 0.08(0.00) & 0.05 & 60481.50(0.07) & -19.65(0.03) & Y\\
WFST-PS240507h & 2024kml & N & 17.30(0.10) & 13:57:20.16 & -1:51:00.46 & 0.17(0.00) & 0.04 & 60453.48(0.12) & -19.96(0.03) & Y\\
WFST-PS240513c & 2024jbb & Y & 19.40(0.01) & 13:38:03.32 & +03:10:26.28 & 0.08(0.00) & 0.02 & 60458.53(0.12) & -19.53(0.03) & Y\\
  \noalign{\smallskip}\hline
\end{tabular}
\ec
\tablecomments{1\textwidth}{\\
$^a$ Redshifts derived from photometric measurements are marked with star symbols.\\
$^b$ SNe firstly discovered by WFST are marked by ``Y".}
\end{table}

Uncertainties exist in early-phase K-corrections for SNe Ia, as they require extrapolation beyond 15 days before maximum light. Furthermore, based on K-correction fitting results from SNoopy\footnote{https://csp.obs.carnegiescience.edu/news-items/snoopy-released}, WFST-PS early-phase SNe Ia show minimal K-corrections, with the highest-redshift objects showing K-corrections $\leq 0.1$ mag. SNe Ia show significant spectral diversity in ultraviolet and blue optical wavelengths at very early phase \citep{2002PASP..114..803N}. Due to the intrinsic variation, standard spectral templates (e.g., \cite{2007ApJ...663.1187H}) cannot reflect early light-curve behavior appropriately. Therefore, K-corrections were not applied for our sample. Figure~\ref{fig:lc-salt2-fit} show the observed light-curves and best-fit SALT2 models for 16 WFST-PS early-phase SNe Ia.

\subsection{WFST-PS EExSNe Ia} \label{subsec:eex}
\begin{figure*}
    \centering
    \includegraphics[width=1\textwidth]{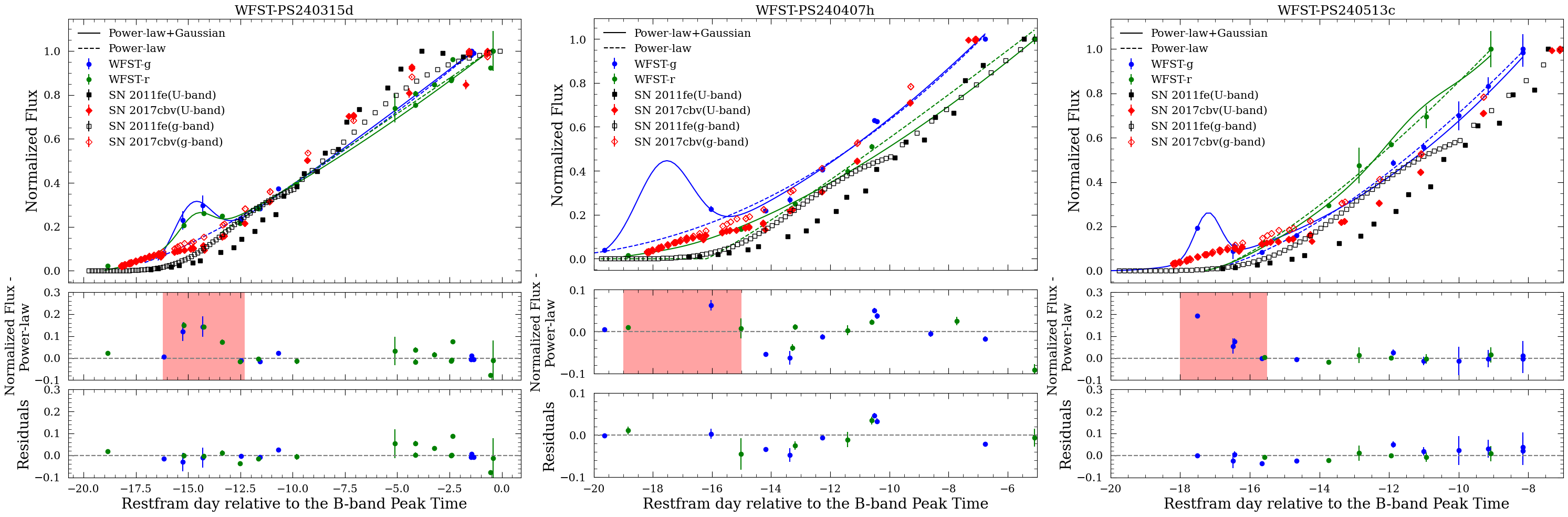}
    \caption{\textbf{Top}: Early light-curves of WFST-PS240315d, WFST-PS240407h, and WFST-PS240513c. The single power-law fits (dashed lines) and power-law + Gaussian fits (solid lines) are shown for each object. $g$-band and $r$-band data with their respective fits are plotted in blue and green. For comparison, the EExSN Ia SN 2017cbv is plotted in red ($U$-band: solid diamonds; $g$-band: open diamonds), while the non-EExSN Ia SN 2011fe is shown in black ($U$-band: solid squares; $g$-band: open squares). \textbf{Middle}: residuals relative to a single power-law fit. \textbf{Bottom panels}: the excess as flux subtracted by the power-law component in the power-law + Gaussian fit, and the corresponding residuals.}
    \label{fig:eex}
\end{figure*}

To adopt a reproducible quantitative approach for identifying early excess flux in early-phase SNe Ia, we performed fitting using a simple power-law function $\alpha(t-t_0)^n$ and conducted simultaneous examinations in both $g$- and $r$-bands following the method in \cite{wu2025commonoriginnormaltype}. For a single band, this requires: (1) two data points with non-zero residuals within this range (with at least one point showing residuals $>2\%$ of peak flux), and (2) all model residuals at these points exceeding their $3\sigma$ uncertainties \citep{Deckers_2022}. In contrast to other WFST-PS early-phase SNe Ia, as shown by Figure~\ref{fig:eex}, the early rising light curves of these three SNe deviate significantly from simple power-law fits, indicating the presence of early-excess emissions.

\textbf{WFST-PS240315d} was discovered at MJD 60380.947, approximately $18.331 \pm 0.064$ days before the rest-frame B-band maximum light as determined by SALT2 fitting. As shown in Figure~\ref{fig:Host}, the supernova is located 4.36$''$ northwest of its host galaxy center. The host's spectroscopic redshift is $z = 0.101$. The SALT2 best fit light-curve show that WFST-PS240315d is a typical normal SN~Ia with a rest-frame $B$-band peak magnitude of $-19.50 \pm 0.02$ and $\Delta m_{15}(B) = 0.97 \pm 0.05$. The photometric behavior of WFST-PS240315d during the first 3 days show significant deviations from that of typical non-EExSNe Ia. As shown in Figure~\ref{fig:eex}, the normalized $g$-band flux reached $0.2$ in the first three days, significantly exceeding that of the EExSNe Ia SN 2017cbv \citep{2019IAUS..339...47H}. This flux excess is over three times brighter than the normal SN Ia 2011fe at the same epoch\citep{2016ApJ...820...67Z}. In the $r$ band, the light-curve behavior is consistent with that observed in the $g$ band in the early phase. As shown in Figure~\ref{fig:eex}, when comparing the single power-law fit to the combined power-law plus Gaussian model, we find significant flux excesses of about 20\% relative to peak brightness in both bands at approximately 15 days before B-band maximum light. The light-curve in the early phase show a distinct "strong-bump" morphology \citep{jiang2018surface}, likely originating from radioactive decay of $^{56}$Ni in the outer ejecta layers.

\textbf{WFST-PS240407h} was discovered at MJD 60399.792, approximately $18.412 \pm 0.056$ days before rest-frame $B$-band maximum as determined by SALT2 fitting. As shown in Figure~\ref{fig:Host}, the supernova is located 1.13$''$ southwest of its host galaxy center. The host's spectroscopic redshift is $z = 0.085$. SNID spectral classification confirms it as a normal SNe Ia with a rest-frame $B$-band peak magnitude of $-19.6 \pm 0.03$ and $\Delta m_{15} = 1.06 \pm 0.07$. The photometric behavior of WFST-PS240407h during the first 2 days show significant deviations from that of typical non-EExSNe Ia. Figure~\ref{fig:eex} reveals that the normalized $g$-band flux reached about 0.1 at first 2 days. In the $g$ band, WFST-PS240407h show early time light-curve behavior similar to SN 2017cbv.

\textbf{WFST-PS240513c} was discovered at MJD 60439.64, approximately $18.89 \pm 0.12$ days before the rest-frame B-band maximum as determined by SALT2 fitting. As shown in Figure~\ref{fig:Host}, the supernova is located 3.75$''$ southwest of its host galaxy center. The host's spectroscopic redshift is $z = 0.078$. SNID spectral classification identifies it as a normal type Ia supernova with a rest-frame B-band peak magnitude of $-19.53 \pm 0.03$ and $\Delta m_{15} = 0.87 \pm 0.15$. The photometric behavior of WFST-PS240513c during the first 2 days show significant deviations from that of typical non-EExSNe Ia. Figure~\ref{fig:eex} show that the normalized flux reached 0.2 at 17 days before maximum in $g$-band, slightly exceeding the early U-band excess observed in SN 2017cbv. Further analysis of the EExSNe Ia sample is presented in Section ~\ref{sec:dis}.

The remaining 13 early-phase SNe Ia were classified as non-EExSNe Ia based on power-law fits in both bands. These non-EExSNe Ia show an average rest-frame B-band peak magnitude of $-19.01 \pm 0.03$ with mean $\Delta m_{15}(B) = 0.98 \pm 0.12$. Further analysis of the non-EExSNe Ia sample is presented in Section~\ref{sec:res}. 
\subsection{Host Galaxies} \label{subsec:Host}
\begin{figure*}
    \centering
    \includegraphics[width=1\textwidth]{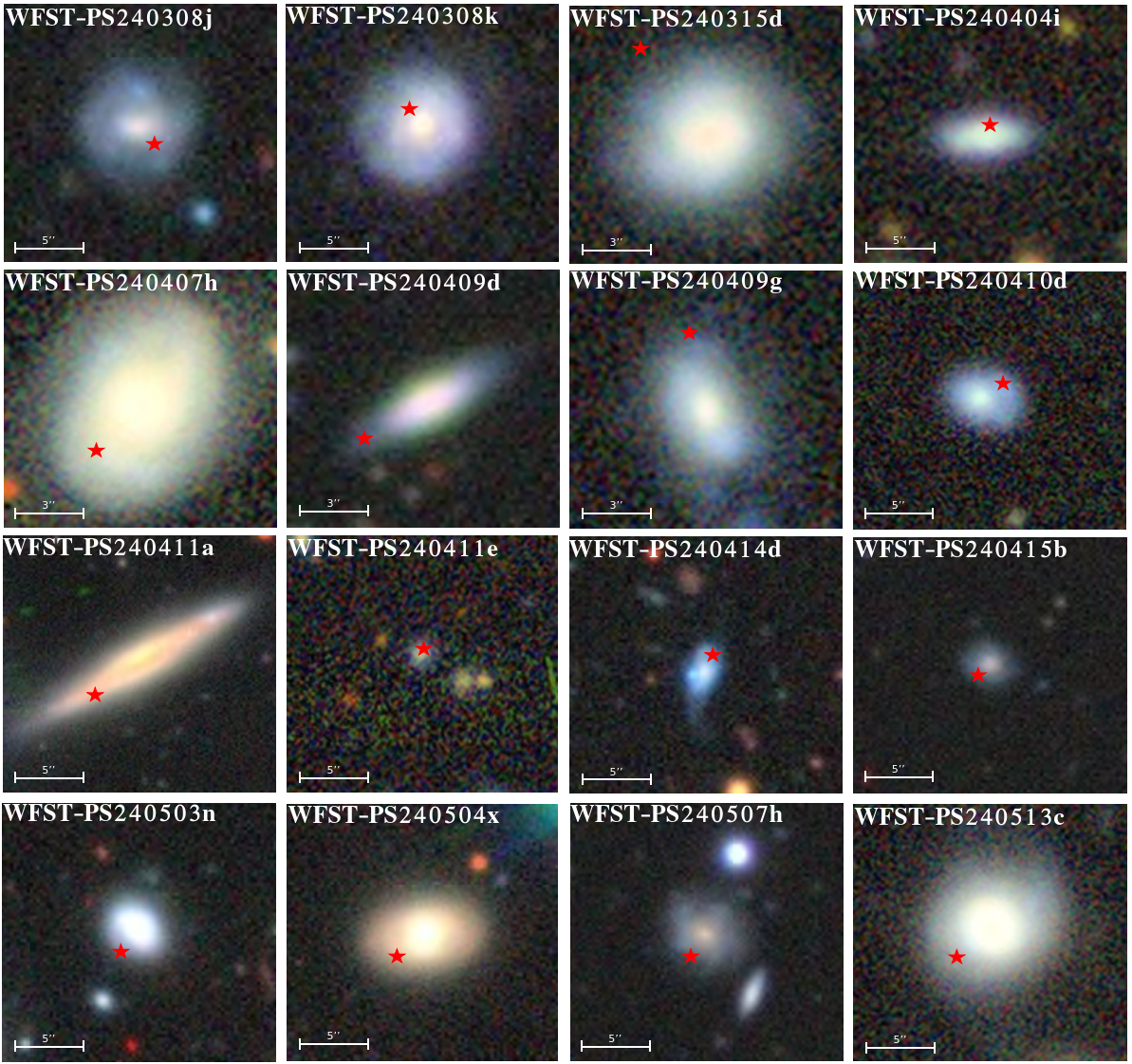}
    \caption{Color images of WFST-PS early-phase SN Ia hosts. The position of each supernova within its host galaxy is marked by a red pentagram. The images were obtained from the DESI DR1 data releases \citep{2025arXiv250314745D}.}
    \label{fig:Host}
\end{figure*}

For each supernova in our sample, we identified its host galaxy using the directional light radius (DLR) method \citep{2006ApJ...648..868S,2016AJ....152..154G} applied to HSC archival images \citep{2021PASJ...73..735T}. Specifically, the galaxy geometrically closest to the SN position was designated as the host, with a selection criterion of DLR $< 10$ \citep{2016AJ....152..154G}. Figure~\ref{fig:Host} shows color images of all early-phase SN Ia host galaxies. According to the NASA IPAC  Extragalactic Database (NED) or Hubble sequence, the host galaxy classifications for our early-phase SNe Ia are listed in Table~\ref{tab:host}. The sample includes spiral (S), elliptical (E), lenticular (S0), and irregular (I) galaxy types. There is no clear host type preference for the SNe sample.

\begin{table}
\begin{center}
\caption[]{ Host Galaxy Information of WFST-PS early phase SNe Ia}\label{tab:host}
 \begin{tabular}{llllllll}
  \hline\noalign{\smallskip}
Name & R.A. & Dec & SFR ($M_\odot~\mathrm{yr}^{-1}$) & Stellar Mass ($M_\odot$) & Offset (kpc)  & $M_g$ & Type$^a$\\
  \hline\noalign{\smallskip}
WFST-PS240308j & 13:21:22.50 & +01:15:03.86 & 1.7905 & 5.9277e+9 & 1.95 & -19.81 & S$^*$\\
WFST-PS240308k & 10:15:43.28 & +04:40:57.55 & 2.0464 & 1.6275e+10 & 2.04 & -19.79 & S$^*$\\
WFST-PS240315d & 13:35:44.83 & +05:19:34.94 & 3.6354 & 5.727e+9 & 8.43 & -20.03 & S\\
WFST-PS240404i & 13:27:20.39 & +02:18:34.21 & 1.8278 & 6.4273e+9 & 0.23 & -19.46 & E$^*$\\
WFST-PS240407h & 14:32:12.23 & +04:59:34.06 &   &   & 1.87 & -21.16 & Sbc\\
WFST-PS240409d & 11:32:47.32 & -2:24:45.99 & 2.5506 & 1.4452e+10 & 3.25 & -19.95 & S$^*$\\
WFST-PS240409g & 13:09:09.76 & +04:54:27.77 & 0.018923 & 4.3816e+7 & 0.43 & -17.90 & S$^*$\\
WFST-PS240410d & 14:09:42.89 & +03:58:14.24 & 1.5708 & 5.2139e+9 & 3.82 & -19.07 & E$^*$\\
WFST-PS240411a & 10:18:28.54 & -1:55:43.81 &  &   & 1.66 & -17.69 & S\\
WFST-PS240411e & 14:27:21.34 & +02:15:44.07 & 0.32671 & 8.0248e+8 & 0.29 & -18.08 & E$^*$\\
WFST-PS240414d & 14:25:09.97 & -1:11:33.69 & 1.467 & 2.0764e+9 & 6.60 & -19.99 & I$^*$\\
WFST-PS240415b & 10:54:18.96 & +01:23:59.30 & 0.90311 & 2.8137e+9 & 5.18 & -18.38 & E$^*$\\
WFST-PS240503n & 13:32:53.36 & -1:27:35.98 & 1.7393 & 7.1602e+9 & 3.03 & -19.95 & E$^*$\\
WFST-PS240504x & 14:49:02.04 & -2:13:09.11 & 2.4019 & 4.3598e+10 & 0.60 & -20.15 & E$^*$\\
WFST-PS240507h & 13:57:20.22 & -1:51:01.28 & 2.2002 & 1.0233e+10 & 3.77 & -19.58 & S$^*$\\
WFST-PS240513c & 13:38:03.10 & +03:10:27.97 & 1.0084 & 7.778e+9 & 5.78 & -19.52 & E\\
  \noalign{\smallskip}\hline
\end{tabular}
\end{center}
\tablecomments{1\textwidth}{\\
$^a$ Hubble classification of galaxies from HyperLeda \citep{2024IAUGA..32P1137M}. The types derived from visual classification are marked with star symbols.}
\end{table}

\begin{figure*}
    \centering
    \includegraphics[width=1\textwidth]{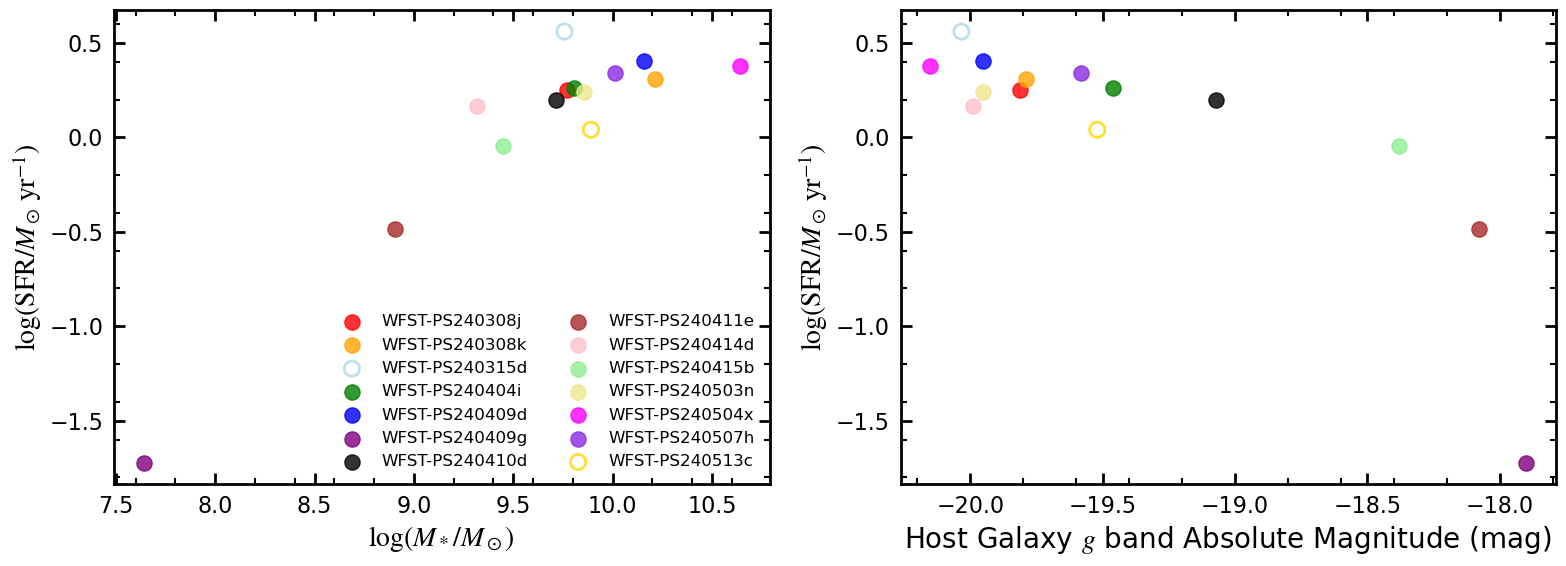}
    \caption{\textbf{left}: Stellar mass versus star formation rate (SFR) distribution for host galaxies of our WFST-PS early-phase SNe Ia sample. The hosts of EExSNe Ia WFST-PS240315d and WFST-PS240513c are denoted by open circles. Those of non-EExSNe Ia are denoted by solid circles. \textbf{Right}: $g$-band absolute magnitude versus star formation rate (SFR) distribution for host galaxies of our WFST-PS early-phase SNe Ia sample.}
    \label{fig:host_info}
\end{figure*}

Using the host galaxy information, we extracted the star formation rates (SFR) and stellar masses ($M_\star$) for each supernova host from the HSC survey data \citep{2015ApJ...801...20T,2018PASJ...70S...9T,2020arXiv200301511N}. Figure~\ref{fig:host_info} compares SFR and $M_\star$ of early-phase SNe Ia in our sample. The host galaxies have a mean stellar mass of $8.76 \times 10^{9}$ $M_\odot$ and a mean SFR of 1.56 $M_\odot$ yr$^{-1}$, with individual masses and DLR values provided in Table~\ref{tab:host}.

\section{Results} \label{sec:res}
\subsection{The Luminosity-Decline Relation}
\begin{figure*}
    \centering
    \includegraphics[width=1\textwidth]{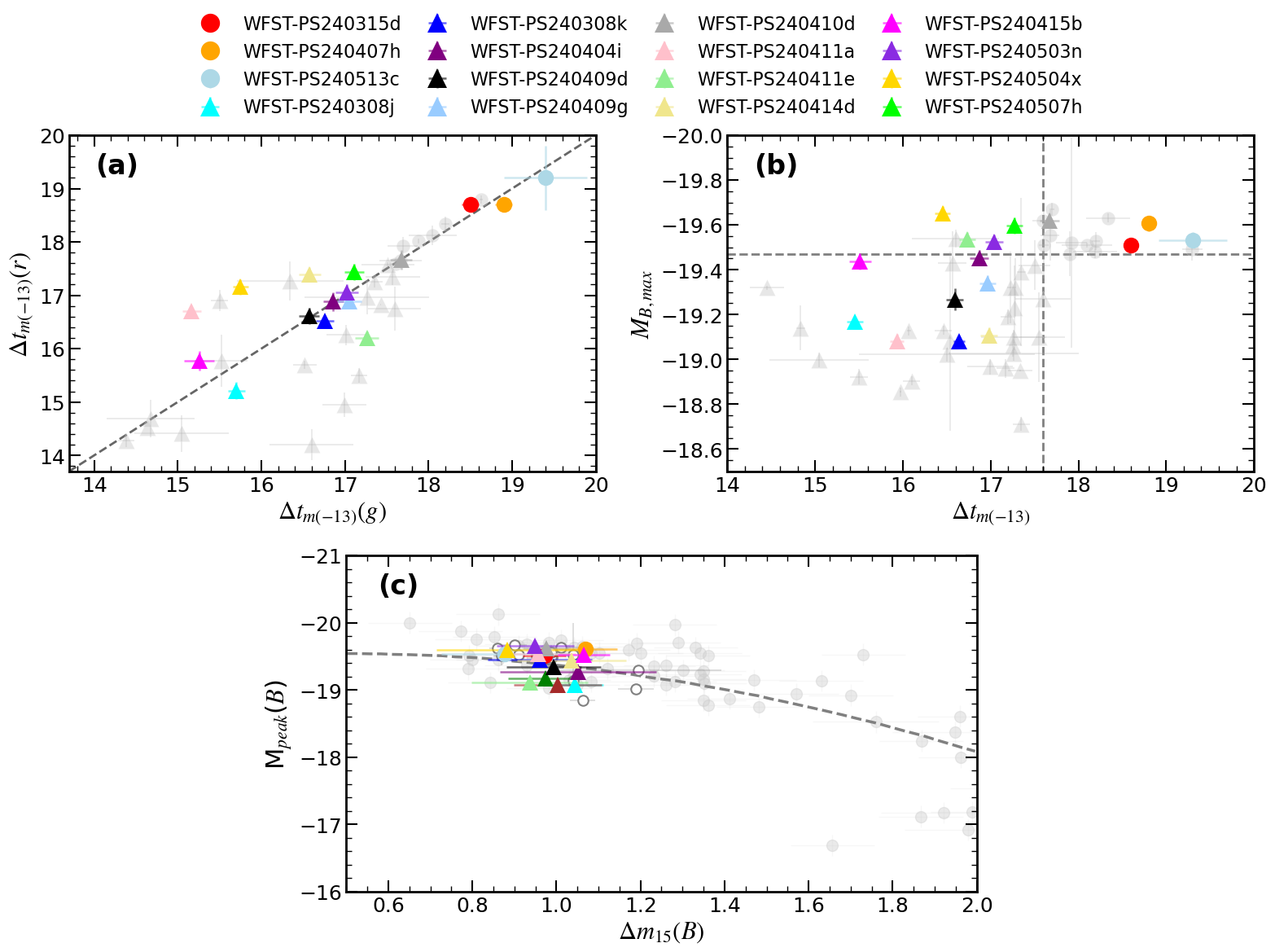}
    \caption{\textbf{Upper left panel (a)}: Distribution of $\Delta t_{m(-13)}$ in the $g$ and $r$ bands. EExSNe Ia and non-EExSNe Ia are shown as circles and triangles, respectively. Data from literature are colored gray. All EExSNe Ia are clustered in the upper right region. Additionally, $\Delta t_{m(-13)}$ values in the $g$ and $r$ bands show relative consistency. \textbf{Upper right panel (b)}: $M_{B,max}$ versus $\Delta t_{m(-13)}$ for early-phase SNe Ia. EExSNe Ia are marked as circles. Gray dashed lines indicate specific thresholds of $\Delta t_{m(-13)}$, $M_{B,max}$, and $\Delta m_{15}(B)$ that separate EExSNe Ia and non-EExSNe Ia samples into distinct quadrants. Notably, all EExSNe Ia are confined to the upper right corner. \textbf{Lower left panel (c)}: $M_{B,max}$ versus $\Delta m_{15}(B)$ for early-phase SNe Ia. Gray open circles denote literature EExSNe Ia. Colored and gray points are from the CfA3 samples in \citet{2009ApJ...700.1097H}. The Phillips relation is highlighted with a gray dashed line.}
    \label{fig:trise}
\end{figure*}

Accurate host galaxy extinction estimates are crucial for establishing the luminosity-decline relation in SNe Ia \citep{1993ApJ...413L.105P}. An empirical approach involves deriving the host extinction corrected absolute $B$-band magnitude by subtracting 3.1 the SALT2 color parameter ($c$) from the Milky Way extinction corrected $M_{B,max}$ \citep{2014A&A...568A..22B}. Figure~\ref{fig:trise} displays the relationship between the c-corrected $M_{B,max}$ and the light-curve decline rate $\Delta m_{15(B)}$ for 16 WFST-PS early-phase SNe Ia. The non-EExSNe Ia in our sample show a correlation between luminosity and decline rate. This suggests that our non-EExSNe Ia are at least photometrically consistent with normal SNe Ia, which agrees with our Superphot+ classification results. For the EExSNe Ia in our sample, similar to other previously reported EExSNe Ia in the literature, our three EExSNe Ia occupy the upper-left region in the Phillips relation, showing brighter peak luminosities and slower decline rates.

\subsection{Rise Time}
To calculate the rise time, the first step is to estimate the explosion epochs for both EExSNe Ia and non-EExSNe Ia. By following \cite{wu2025commonoriginnormaltype}, we applied $\Delta t_{m(-13)}$ as the indicator of rise time. In our calculation of rise times ($\Delta t_{m(-13)}$) for the early-phase SNe Ia sample, we adopt $t_{m(-13)}$ as the zero point, considering that SNe Ia show negligible rise time before reaching $-13$ magnitude and remain unaffected by any early excess (EEx) mechanisms. We performed two distinct light-curve model fits for the WFST-PS early-phase SNe Ia sample \citep{2024ApJ...962...17W}:

A single power-law model:
\begin{equation}
    f(t) = A_{\text{pl}}(t - t_{m(-13)})^{\alpha}
\end{equation}
    
A composite model:
\begin{equation}
    f(t) = A_{\text{pl}}(t - t_{m(-13)})^{\alpha} + \left(\frac{A_G}{\sigma\sqrt{2\pi}}\right)e^{-(t-\mu)^2/2\sigma^2}
\end{equation}

where $A_{\text{pl}}$ and $\alpha$ represent the amplitude and exponent of the power-law component. $A_G$, $\mu$, and $\sigma$ correspond to the amplitude, center point, and width of the Gaussian component. $t_{m(-13)}$ denotes the time of $B$-band maximum minus 13 days.

\begin{figure*}
    \centering
    \includegraphics[width=1\textwidth]{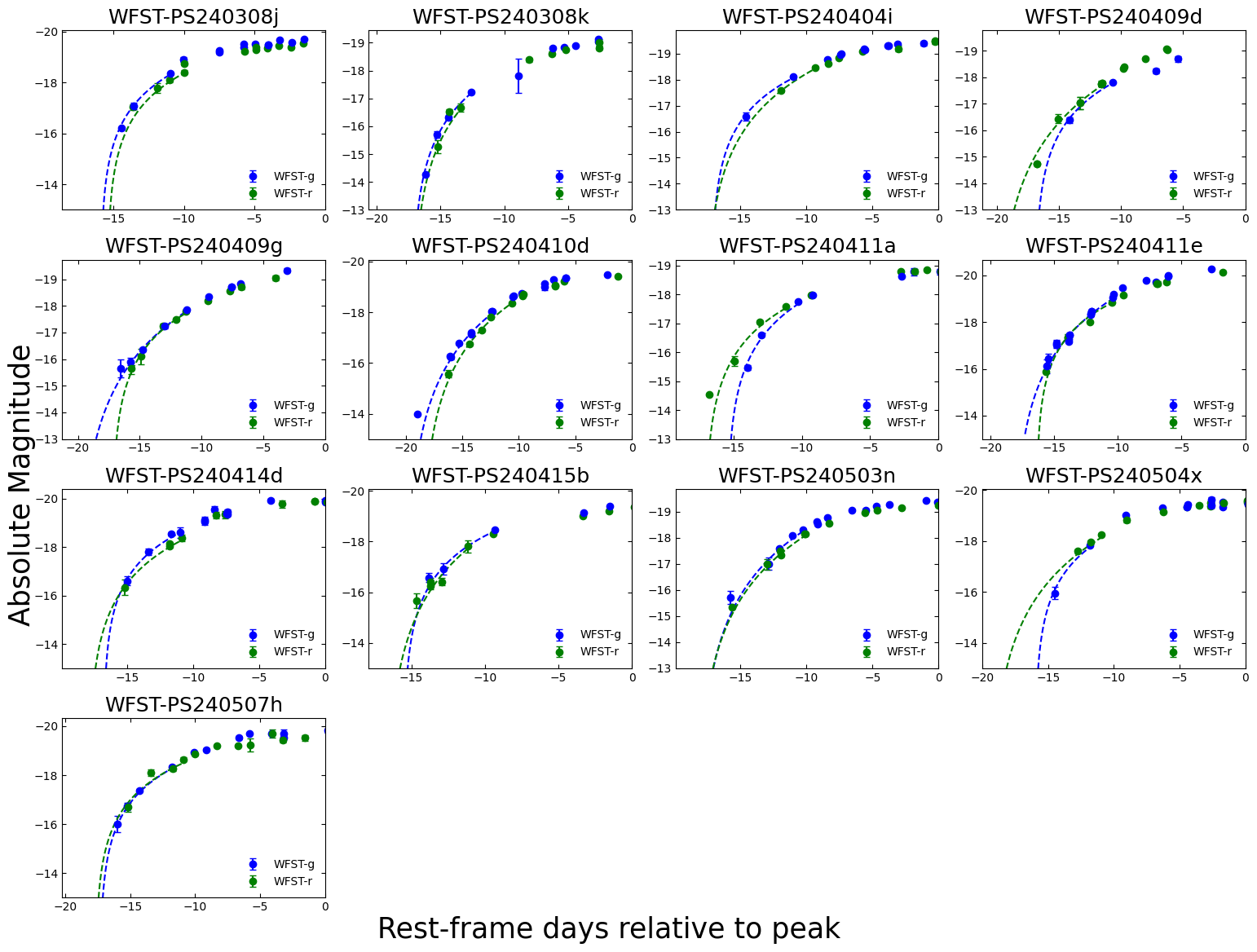}
    \caption{Early light-curve fitting of non-EExSNe Ia before the $B$-band peak time using a single power law. Blue and green represent the $g$- and $r$-band data and fits, respectively. The fitting criteria require normalized flux $\leq 0.4$; where insufficient data are available, the threshold is relaxed to normalized flux $\leq 0.5$. 
    \label{fig:rise-lc}}
\end{figure*}

We first performed least-squares fitting using the lmfit package and evaluated the goodness-of-fit through the Bayesian Information Criterion (BIC). We focused on three key parameters: $t_{m(-13)}$, and the power-law indices $\alpha_g$ and $\alpha_r$ in $g$- and $r$-bands, respectively. The prior parameter range for the power-law indices was constrained to $1 < \alpha < 3$. We then approximated the posterior distributions using Bayes' theorem with affine-invariant MCMC sampling. The implementation used the emcee package \citep{foreman2013emcee}, running 100 chains until convergence or a maximum of 3 million steps per chain, with the convergence criterion set at $n_{\text{steps}} > 100\tau$ (where $\tau$ represents the autocorrelation length; \citealt{2020ApJ...902...47M}). For the EExSNe Ia sample WFST-PS240315d, WFST-PS240407h and WFST-PS240513c, as shown in Figure~\ref{fig:eex}, the fitting residuals approach zero when including the Gaussian component, consistent with the detected flux excess in these sources. For the remaining non-EExSNe Ia, Table~\ref{tab:sample} presents their $g$- and $r$-band $t_{m(-13)}$ values derived from single power-law fitting, with the corresponding light-curve fits displayed in Figure~\ref{fig:rise-lc}.

Using the derived $t_{m(-13)}$ values, we analyzed the sample's rise times $\Delta t_{m(-13)}$. Since $t_{m(-13)}$ is measured relative to $T_{B,\text{max}}$ (which itself carries measurement uncertainties), the $\Delta t_{m(-13)}$ estimation must account for errors in both $t_{m(-13)}$ and $T_{B,\text{max}}$ \citep{2020ApJ...902...47M}. For the EExSNe Ia, the rise times in $g-$band for WFST-PS240315d, WFST-PS240407h and WFST-PS240513c are: $18.60 \pm 0.07$, $18.79 \pm 0.07$ and $19.29 \pm 0.39$, respectively. As shown in panel (c) of Figure~\ref{fig:trise}, these EExSNe Ia occupy the upper-left region in the Phillips relation, indicating both brighter peak luminosities and longer rise times compared to normal SNe Ia. This result is consistent with the finding of \cite{wu2025commonoriginnormaltype}. For non-EExSNe Ia, the measured average rise time for the entire sample is 16.89 days. Based on the sample standard deviation, the dispersion of the mean is approximately 0.052 day. For the $\alpha_g$ index, the sample average is $\sim\!2$, with a typical dispersion of $0.1$. The sample show $\alpha_g$ values as low as $\sim\!1.5$ or as high as $\sim\!2.5$, indicating a significant variation in $\alpha_g$. For the $\alpha_r$ index, the sample average is $\sim\!1.9$, with a typical dispersion of $0.1$. In both the $g$ and $r$ filters, the mean rise power-law index during the initial evolutionary phase of the SNe approaches $2$, consistent with predictions from the expanding fireball model \citep{1999AJ....118.2675R}.

\subsection{Color}
\begin{figure*}
    \centering
    \includegraphics[width=1\textwidth]{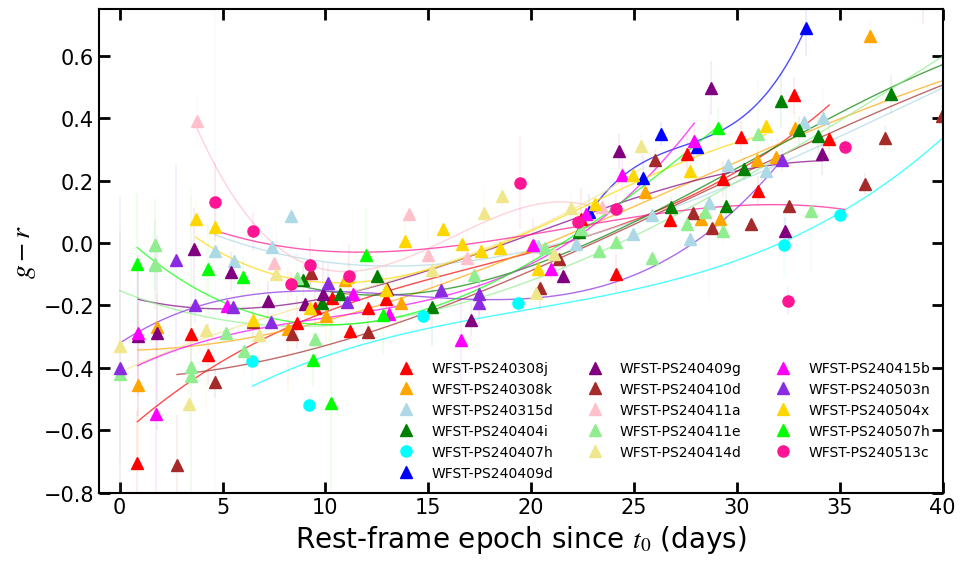}
    \caption{ $g-r$ color evolution for 16 WFST-PS early-phase SNe Ia. Colors are corrected for reddening and fitted with solid curves. EExSNe Ia and non-EExSNe Ia are represented by circles and triangles, respectively.}
    \label{fig:color}
\end{figure*}

Thanks to the observational strategy of the WFST-PS transient survey, our early-phase SNe Ia obtained color information more than 15 days before $B$-band maximum. To reduce uncertainties, we averaged observations taken on the same night and selected detections with significance above $3\sigma$. We then calculated the daily $g-r$ color indices from the $g$- and $r$-band measurements. The photometric data were corrected for time dilation effects and Galactic extinction. Following \citet{2020ApJ...902...48B}, we performed weighted least-squares linear fits to the early phase $g-r$ data (within $\sim\!10$ days before $B$-band maximum) to determine the color evolution slope $\Delta(g-r)/\Delta t$ for each supernova, quantifying their color change characteristics. Figure~\ref{fig:color} displays the $g-r$ color distribution of 16 early-phase SNe Ia discovered by WFST-PS during their pre-maximum phase relative to $B$-band peak.

\begin{figure*}
    \centering
    \includegraphics[width=1\textwidth]{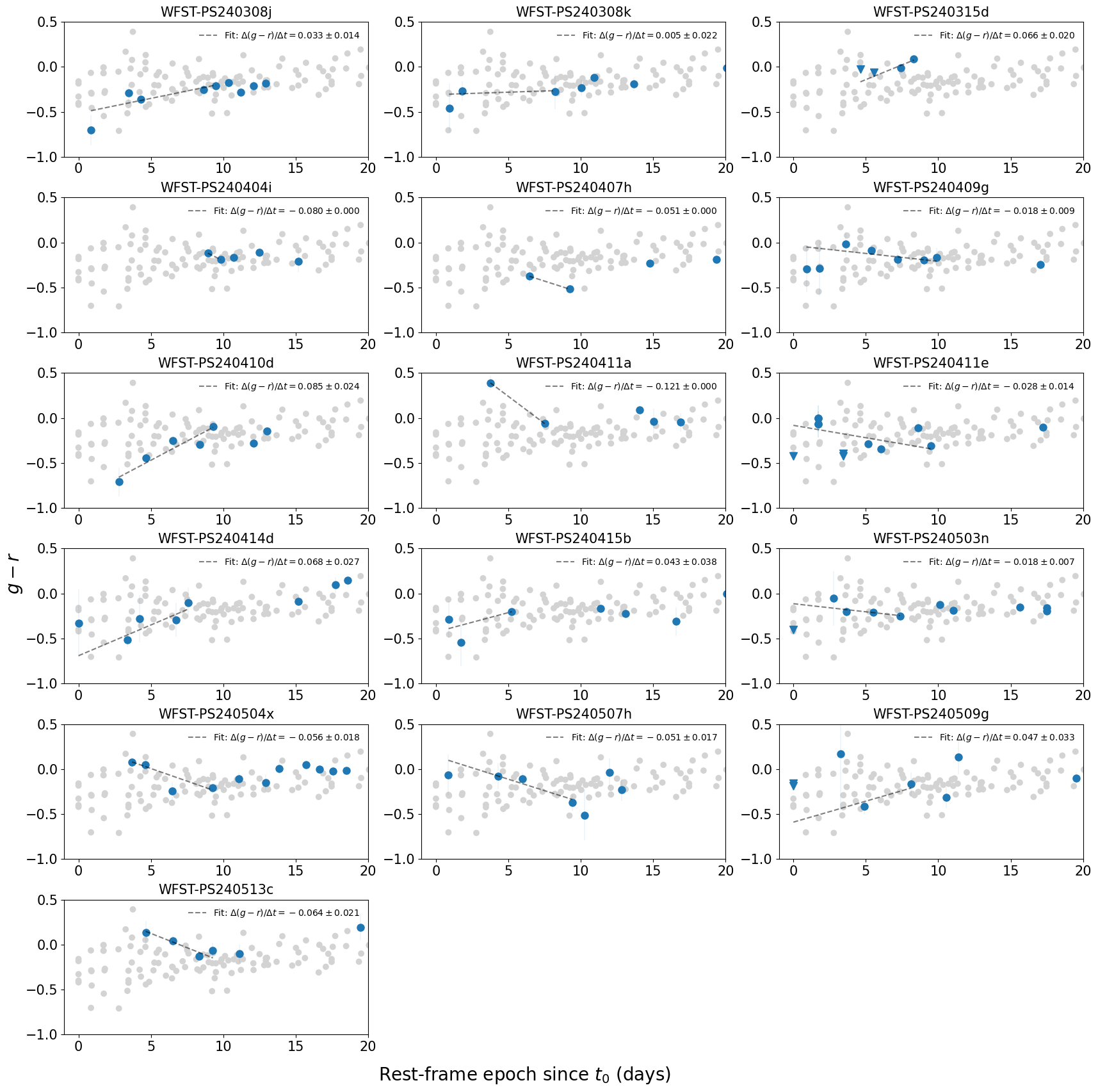}
    \caption{Evolution of $g - r$ colors for 16 WFST-PS early-phase SNe Ia} discovered by WFST-PS within 20 days from $t_{0}$. Grey points mark the colors of the full sample for comparison. The black dashed line in each panel is a weighted least-square linear fit to colors in the first 10 days for events with at least two data points in this time window.
    \label{fig:color-fit}
\end{figure*}

For non-EExSNe Ia, the $g - r$ color distribution appears relatively uniform starting from approximately 10 days before $B-$band maximum, with colors concentrated around $g - r \sim -0.1$ mag and showing a scatter of 0.15 mag. In contrast, the early colors show greater dispersion, with a $g - r$ scatter of 1.25 mag. Figure~\ref{fig:color-fit} presents the $\Delta(g - r)/\Delta t$ fitting results, while Table~\ref{tab:sample} lists the specific values for 16 SNe. As shown in Figure~\ref{fig:color-fit}, some SNe display a red to blue transition, while others show nearly flat evolution. The results indicate that 10 objects appear blue-dominated and 7 appear red-dominated. The slope $\Delta(g - r)/\Delta t$ distribution ranges from -0.121 to 0.085 mag/day, forming a continuous distribution that does not align with the bimodal classification proposed by \cite{stritzinger2018red}. Furthermore, comparing our sample with the 12 SNe Ia from \cite{stritzinger2018red}that have either $g-$ and $r-$band photometry results, we find that their $B - V$ based classification into "red" and "blue" groups does not show clear separation in early $g - r$ colors. This supports the view that early color evolution of SNe Ia may be more homogeneous in $g-$ and $r-$band filters \citep{2020ApJ...902...48B}.

\begin{figure*}
    \centering
    \includegraphics[width=1\textwidth]{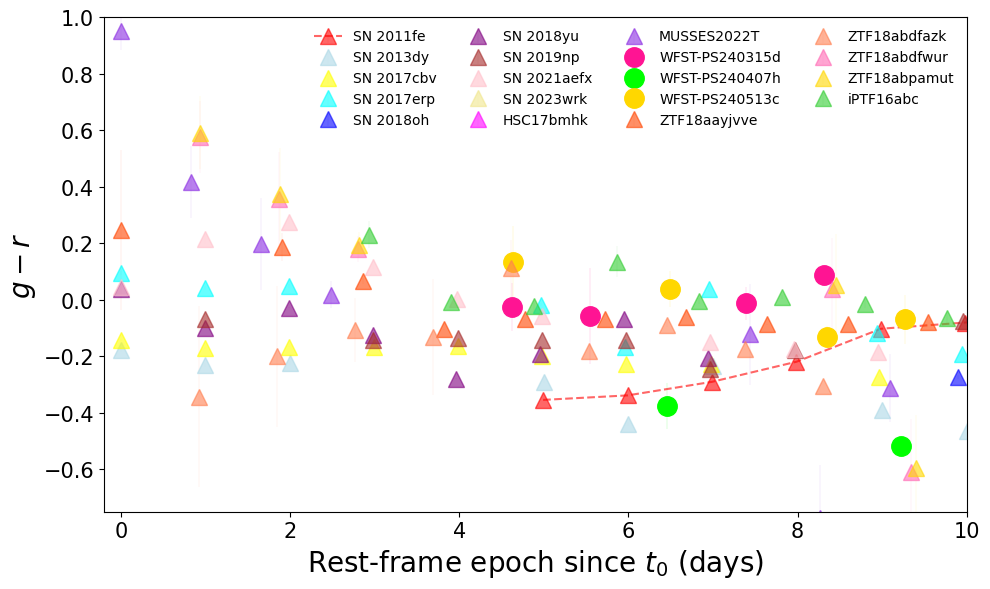}
    \caption{Evolution of $g - r$ colors for the 18 EExSNe Ia within 10 days from first-light $t_{0}$. WFST-PS240315d, WFST-PS240407h, and WFST-PS240513c are represented by circular symbols. EExSNe Ia from the literature are denoted by triangles. The non-EExSNe Ia SN 2011fe is included for comparison. 
    \label{fig:color-eex}}
\end{figure*}

For EExSNe Ia, the $\Delta(g - r)/\Delta t$ values of WFST-PS240315d, WFST-PS240407h, and WFST-PS240513c are approximately 0.066, $-0.051$, and $-0.064$, respectively. Compared to other EExSNe Ia in literature (Figure~\ref{fig:color-eex}), WFST-PS240315d and WFST-PS240513c show color evolution similar to that of iPTF16abc at comparable phases. WFST-PS240407h shows a bluer evolution, comparable to ZTF18aayjvve. A detailed discussion of the color analysis for samples with UV-optical data is provided in the section~\ref{sec:dis}.

\section{Discussion} \label{sec:dis}
\subsection{Ejecta-companion Interaction} 
\begin{figure*}
    \centering
    \includegraphics[width=1\textwidth]{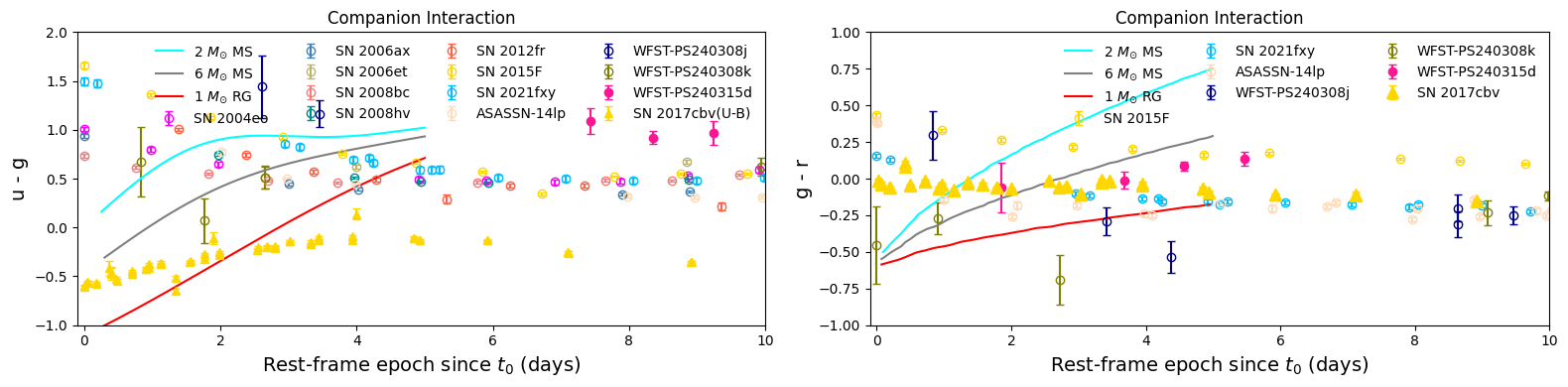}
    \caption{Comparison of our WFST-PS sample to SN ejecta-companion models. Predicted colors are shown only in the first 5 days since $t_{0}$ when the emission from the SN ejecta-companion interaction is expected to be dominant. We model three different progenitor scenarios: a RG companion at $a = 2 \times 10^{13} cm$, a 6 $M_{\odot}$ main-sequence star ($a = 2 \times 10^{12} cm$) and a 2 $M_{\odot}$ main-sequence star ($a = 5 \times 10^{11} cm$). For all color curves shown here, the viewing angle is $\theta = 0^\circ$. EExSNe Ia and non-EExSNe Ia from literature are denoted by solid and open symbols, respectively.
    \label{fig:CSM}}
\end{figure*}

For companion interaction modeling, we employ the analytic formulae from \cite{2010ApJ...708.1025K} (Equations 22 and 25) to calculate blackbody radiation across different binary separations ($a$). Assuming perfect alignment between the white dwarf, companion star, and observer during explosion, we further estimate flux evolution and corresponding color changes within 5 days after first light. We consider three companion systems: 1 $M_{\odot}$ red giant ($a = 2 \times 10^{13}$ cm); 6 $M_{\odot}$ main-sequence star ($a = 2 \times 10^{12}$ cm); 2 $M_{\odot}$ main-sequence star ($a = 5 \times 10^{11}$ cm). The calculation assumes the ejecta velocity of $10^{4}$ km s$^{-1}$ and effective opacity of 0.2 cm$^{2}$g$^{-1}$. Figure~\ref{fig:CSM} compares our sample with the supernova ejecta companion interaction models proposed by \cite{2010ApJ...708.1025K}.

All models predict similarly blue colors at first light ($g - r \approx -0.3$ mag), followed by gradual reddening as temperatures decline. This blue to red evolution manifests as $\Delta(g - r)/ \Delta t \approx 0.3$ mag day$^{-1}$, though the color evolution rate diminishes when companions are red giants or higher mass main sequence stars. \cite{2010ApJ...708.1025K} estimates that only $10\%$ of companion interaction cases would be observationally detectable. While we cannot exclude the possibility of companions existing for every supernova, \cite{2019IAUS..339...47H} overpredicted UV flux in their best fit model for SN 2017cbv, and similar discrepancies emerged in \citet{2020ApJ...902...48B} optical color fitting for SN 2019yvq. Furthermore, the model expected hydrogen or helium emission lines have never been detected in these EExSNe Ia \citep{2018ApJ...863...24S,2020ApJ...900L..27S}.

Interaction models typically predict excess emission persisting 2-3 days with enhanced $UV$-band dominance, as shown in Figure~\ref{fig:CSM}. Additional radiation sources from internal $^{56}$Ni contributions may influence light-curves and color evolution, particularly in close binary systems with small separations. For systems with $a= 5 \times 10^{11}$ cm (2 $M_{\odot}$ main-sequence star), the expected radiative thermal luminosity from $^{56}$Ni contributions could surpass the radiative output from shock thermalization. Given the anticipated slower color evolution associated with $^{56}$Ni contributions, optical emission may become particularly contaminated by this additional energy source. Conversely, UV output likely remains minimally affected by such contributions. This dichotomy could complicate interpretations of UV-optical behavior within companion interaction scenarios.

\subsection{Surface $^{56}$Ni Decay} \label{subsec:tables}
Minor modifications to the $^{56}$Ni distribution in the outer layers of SNe Ia can induce significant differences in their light-curve rise phases. A shallow $^{56}$Ni distribution produces a steeper rise, while a deeper distribution corresponds to a prolonged dark phase and gradual rise \citep{2020A&A...634A..37M}. Shallow heating results bluer radiation in early time, making color evolution an effective diagnostic tool. Models with stronger mixing yield characteristically early bluer colors and relatively flat evolutionary curves. \cite{2020A&A...634A..37M} employed the radiative transfer code TURTLS \citep{2018A&A...614A.115M} to compute mixed models. Their light-curve grid was constructed by varying four key parameters: $^{56}$Ni mass (0.4, 0.6, 0.8 $M_{\odot}$), density profile morphology (broken power-law or exponential), kinetic energy, and $^{56}$Ni mixing extent. 

\begin{figure*}
    \centering
    \includegraphics[width=1\textwidth]{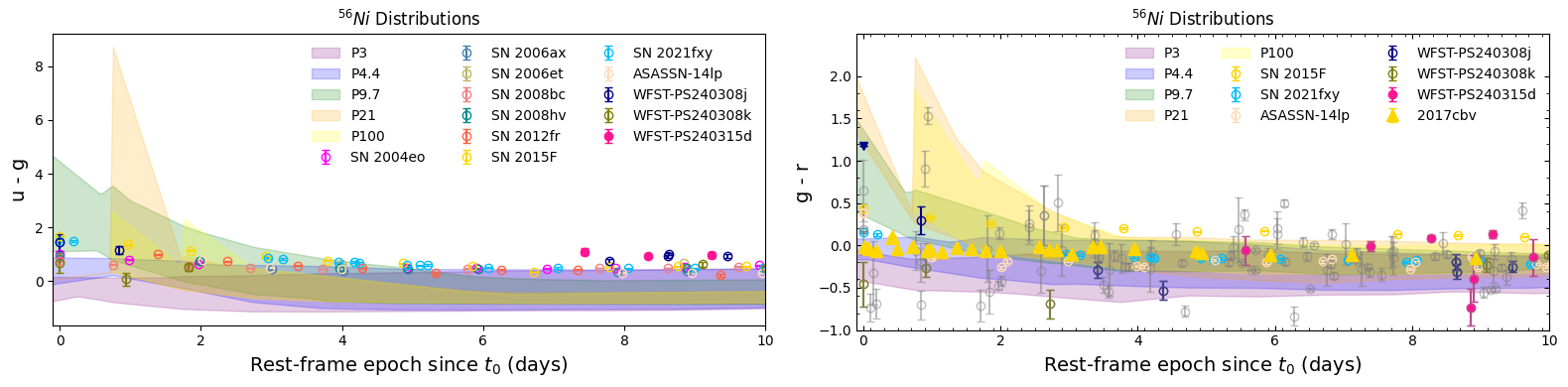}
    \caption{Comparison of our WFST-PS sample to $^{56}$Ni mixing models. Non-EExSNe Ia without $u$-band data are represented by gray points in the right panel. For each mixing model, the shaded area represents color variations for different density profile shapes and kinetic energies (see text for details).
    \label{fig:56Ni_mixing}}
\end{figure*}

Figure~\ref{fig:56Ni_mixing} shows the $u\!-\!g$ and $g\!-\!r$ color evolution of the WFST-PS sample, supplemented with additional SNe for comparative analysis. The observational data reveal significant dynamic variations in color evolution during the first 5 days. Model fitting results are consistent with \cite{2020ApJ...902...48B}, supporting a strong $^{56}$Ni mixing scenario. In contrast, the highly stratified ``P100'' and ``P21'' models are disfavored due to statistically significant discrepancies with the observed early-phase light-curve behavior. However, the best-matching models from \cite{2020A&A...634A..37M} show discrepancies in $u$-band light-curves compared to observations. A representative example is their optimal fit to the gold-standard SN 2012fr, which adopts a model with 0.6 $M_{\odot}$ of $^{56}$Ni, moderate kinetic energy ($1.40$), and intermediate mixing ($9.7$). Even near maximum light, this model fails to reproduce the observed $u$-band light-curve. This suggests that single-component $^{56}$Ni mixing cannot explain early phase observations of SNe Ia. \cite{2020A&A...634A..37M} models suggest that about $22\%$ of SNe Ia require early flux excesses, yet all corresponding models show smooth early rises. This implies that monotonic increases in outer layer $^{56}$Ni abundance cannot produce the observed early flux excesses, necessitating additional radiation sources not accounted for in current models. 

The pulsational delayed detonation (PDDEL) mechanism produces blue early colors and significant luminosity without relying on strong mixing of $^{56}$Ni heating in the ejecta \citep{2014MNRAS.441..532D}. The pulsational effect only influences the outer ejecta layers, thus primarily affecting early phase evolution. The PDDEL model indicated that enhanced outer ejecta temperatures during initial phases produce observable effects on luminosity and colors within days, making SNe both brighter and bluer. The PDDEL4m ($M_{\text{mix}} = 0.25\,M_\odot$) model calculated by \citet{2020ApJ...902...48B} show bluer color indices and flatter evolution curves compared to the non-pulsational delayed detonation model DDC15m ($M_{\text{mix}} = 0.25\,M_\odot$ and a Gaussian smoothing with a characteristic width of $300\,\text{km}\,\text{s}^{-1}$). However, the predicted red to blue transition during the first three days in this model makes it challenging to reproduce the flattest portion of the observed distribution. Furthermore, as shown in Figure~\ref{fig:56Ni_mixing}, the $u - g$ color index difference is more pronounced than $g - r$ in early phases. This further show the greater sensitivity of the $u$-band compared to other optical bands, suggesting that studying $u - g$ color evolution could provide clearer discriminative evidence between different mechanisms.

\subsection{Double-detonation Scenario}

\begin{figure*}
    \centering
    \includegraphics[width=1\textwidth]{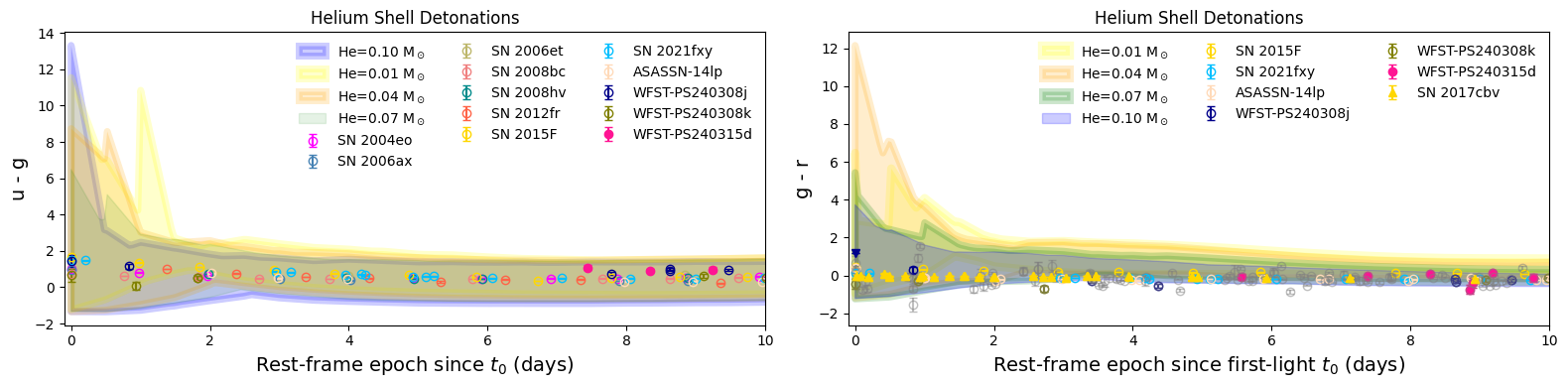}
    \caption{Comparison of our WFST-PS sample to double detonation model. Models from \cite{2021MNRAS.502.3533M} have fixed carbon-oxygen mass at \( 1.0\,M_\odot \) and helium masses varying in the range of \( 0.01\text{-}0.10\,M_\odot \).
    \label{fig:He}}
\end{figure*}

In double detonation models, most studies find that helium shell burning reaches nuclear statistical equilibrium (NSE), producing substantial amounts of $^{56}$Ni and other iron group elements (IGEs). Figure~\ref{fig:He} compares our sample with the supernova double detonation models proposed by \cite{2021MNRAS.502.3533M}. To maximize the coverage of the parameter space for the model's fit to the observed data, we fixed the core mass \( M_{\text{CO}} \) at \( 1.0\,M_\odot \) and varied the helium shell mass \( M_{\text{He}} \) in the range of \( 0.01\text{-}0.10\,M_\odot \). The dominant products of shell burning include IGEs (\( ^{52}\text{Fe} \), \( ^{56}\text{Ni} \), \( ^{48}\text{Cr} \)) and IMEs (\( ^{32}\text{S} \), \( ^{44}\text{Ti} \)). As shown in Figure~\ref{fig:He}, the early color bump morphology in this model is primarily driven by the shell material. Models with IGE rich shells produce a color bump within a few days after the explosion, whereas models without IGE show a relatively flat and blue-color evolution. When the helium shell mass \( M_{\text{He}} \) is \( 0.10\,M_\odot \), the model provides a good coverage of the color evolution range observed in the sample. By individually examining the \( g - r \) color evolution of each supernova in Figure~\ref{fig:color-fit}, we identified a blue-red-blue color bump in WFST-PS240409g, WFST-PS240411e, and WFST-PS240503n, which aligns with the predicted range of the model when the dominant shell material is \( ^{56}\text{Ni} \). However, best fit results reveal that SN 2017cbv $U$-band bump appears more pronounced in models than in observations, with faster post peak decline rates in blue bands ($U$, $B$, and $g$) compared to actual data 3 weeks after explosion. For SN 2019yvq, the model predicts greater $g$-band decline amplitude post-bump peak than observed. Even when employing different shell compositions (e.g., $^{52}$Fe or $^{48}$Cr dominated models), simultaneous matching of $g$- and $r$-band bump features remains elusive. For blue targets, model spectra at maximum light generally show broader line features and stronger flux suppression compared to observations. Regardless of shell composition, no models can simultaneously match both the early light-curves and maximum spectra of these blue targets. 

Improved models aligned with current observational signatures are required to better explain the early light-curve evolution and color behavior of both EExSNe Ia and non-EExSNe Ia. One plausible alternative invokes a thin helium shell double detonation model \citep{wu2025commonoriginnormaltype,piro2025earlyemissiondoubledetonation}, where the secondary detonation arises from an off center helium ignition during the first detonation phase. This mechanism could produce a population of early-phase SNe Ia with smooth color evolution and early excesses flux, modulated by viewing angle effects. In this model, line blanketing from shell material tends to redden colors, while temperature effects from varying $^{56}$Ni masses counteract this trend. Although thicker shells theoretically enhance line blanketing, increased shell mass in this model simultaneously elevates $^{56}$Ni production for a fixed core mass, ultimately maintaining bluer colors via sustained higher ejecta temperatures. Future numerical simulations of this theoretical framework will clarify whether it can unify the diverse origins of SNe Ia. 

\section{Conclusion} \label{sec:con}
In this paper, we present 16 early-phase SNe Ia at redshifts from 0.018 to 0.165 discovered by the WFST-PS in 2024. Three SNe Ia, WFST-PS240315d, WFST-PS240407h, and WFST-PS240513c with early-excess emission features have relatively high peak luminosities and long rise time compared to other 13 non-EExSNe Ia. Generally $c$ and $x_1$ values of all 16 SNe Ia are consistent with those for cosmology-used SNe Ia.

Based on the systematic investigation of color evolution of the WFST-PS sample, we find a large $g-r$ scatter in approximately the first 10 days of SN explosions. However, with limited $u$-band data of the WFST-PS sample, we find that $u-g$ color evolves even more dramatically in the first five days. Comparative analyses of $u-g$ color evolution among models, such as ejecta-companion interaction, models incorporating $^{56}$Ni mixing in outer ejecta, and double detonation with varying carbon-oxygen core masses suggest that current models require refinement to reconcile with observational signatures of SNe Ia in the early photometric behavior. Given the unique $u$-band survey capability of WFST, the ongoing WFST DH$ugr$ project will enable systematic analysis of NUV-optical color evolution, stringently constraining explosion mechanisms and progenitor systems.

\begin{acknowledgements}
This work was supported by the National Natural Science Foundation of China (Grant No. 12393811), the National Key R\&D Program of China (Grant No. 2023YFA1608100), and the Strategic Priority Research Program of the Chinese Academy of Science (Grant No. XDB0550300). J.J. acknowledges support from the Japan Society for the Promotion of Science (JSPS) KAKENHI grants JP22K14069. K.M. acknowledges support from the JSPS KAKENHI grant JP24KK0070 and 24H01810. K.M. and H.K. acknowledge support from the JSPS bilateral JPJSBP120229923. L.G. acknowledges financial support from AGAUR, CSIC, MCIN and AEI 10.13039/501100011033 under projects PID2023-151307NB-I00, PIE 20215AT016, CEX2020-001058-M, ILINK23001, COOPB2304, and 2021-SGR-01270.\\
The Wide Field Survey Telescope (WFST) is a joint facility of the University of Science and Technology of China and Purple Mountain Observatory. This work has made use of the open-source Python packages Astropy \citep{robitaille2013astropy}, emcee \citep{foreman2013emcee}, lmfit \citep{newville2016lmfit}, scipy \citep{virtanen2020scipy}, SALT2 \citep{guy2007salt2} and SNID \citep{2007ApJ...666.1024B}. The figures in this article were created using Matplotlib \citep{hunter2007matplotlib}.
\end{acknowledgements}

\label{lastpage}
\bibliographystyle{raa}
\bibliography{sample7}
\end{document}